\documentclass[structabstract,rnote]{aa}
\usepackage{txfonts}
\usepackage{graphicx}
\usepackage{natbib}

\bibpunct{(}{)}{;}{a}{}{,} 

\def\phx{{\texttt{PHOENIX}}}

\begin{document}

\title{Near-infrared light curves of type Ia supernovae} 
\author{D. Jack\inst{1}
  \and P. H. Hauschildt\inst{1} 
   \and E. Baron\inst{1,2}} 

\institute{Hamburger Sternwarte, Gojenbergsweg 112, 21029 Hamburg, Germany\\
  e-mail: djack@hs.uni-hamburg.de; yeti@hs.uni-hamburg.de
  \and Homer L. Dodge Department of Physics and Astronomy, University of Oklahoma, 440 W Brooks, Rm 100, Norman, OK 73019-2061 USA\\
  e-mail: baron@ou.edu}

\date{Received 17 May 2011 /
     Accepted 3 January 2012}

    \abstract {} 
{With our time-dependent model atmosphere code \phx,
      our goal is to simulate light curves and spectra of
      hydrodynamical models of all types of supernovae. In this work,
      we simulate near-infrared light curves of SNe Ia and confirm the
      cause of the secondary maximum.}  
{We apply a simple energy
      solver to compute the evolution of an SN Ia envelope during the free expansion phase. 
      Included in the solver are energy changes due
      to expansion, the energy deposition of $\gamma$-rays and
      interaction of radiation with the material.}
{We computed
      theoretical light curves of several SN Ia hydrodynamical models
      in the I, J, H, and K bands and compared them to the observed SN
      Ia light curves of SN 1999ee and SN 2002bo.  By changing a line
      scattering parameter in time, we obtained quite reasonable fits
      to the observed near-infrared light curves. This is a strong
      hint that detailed NLTE effects in IR lines have to be modeled,
      which will be a future focus of our work.}  
{We found that IR
      line scattering is very important for the near-infrared SN Ia
      light curve modeling. In addition, the recombination of Fe III
      to Fe II and of Co III to Co II is responsible for the secondary
      maximum in the near-infrared bands. For future work the
      consideration of NLTE for all lines (including the IR
      subordinate lines) will be crucial.}


\maketitle

\section{Introduction}

We focus on the near-infrared light curves of type Ia supernovae (SNe Ia). The
light curves in the near-infrared are of particular interest, because
they have been claimed to be near standard candles at the time of
B-band maximum in the IR \citep{KPS04}. In addition, many, but not all
SNe~Ia exhibit 
a secondary maximum in the IR bands. The secondary maximum was first
noted by \citet{elias81}. \citet{hofdd+mol95} explained the secondary maximum
as due to the expansion of the IR pseudo-photosphere, whereas
\citet{suntz96} suggested that it was due to a global shift of
radiation from blue to red. \citet{PE00} suggested that the transition
from Fe III to Fe II as the dominant ionization stage was important
although they implied that the secondary maximum occurs when the
photosphere has receded into the non-radioactive center. 
\citet{kasen06b} performed
a detailed study of SN Ia light curves in the near-infrared and
obtained reasonable fits to observations.  He also finds that the secondary
maximum is an effect of the ionization evolution of iron group
elements in the expanding envelope.  We apply our time-dependent model
atmosphere code \phx\ and investigate SNe Ia by comparing model light
curves in the near-infrared to observed light curves.  We will also
focus on the origin of
the secondary maximum.

We present model light curves of SNe Ia in the
near-infrared.  In Sect. \ref{sec:straw_man} we present the
methods we used in some detail.
In Sect. \ref{sec:line_scat} we
present light curves of SNe~Ia for different hydrodynamical models
in the I, J, H, and K bands.  
We consider a parametrized IR line scattering in the solution of the
radiative transfer to improve our fits to observed light
curves. An investigation of the secondary maximum that has been
observed in near-infrared light curves is discussed in a final section.

\section{Methods}
\label{sec:straw_man}

We use our time dependent model atmosphere code \phx, version 16, to
compute model light curves of type Ia supernovae.  We use the time
dependent extension and the method as described in \citet{jack09,jack11}.
We calculate SN Ia model light curves for different hydrodynamic (explosion)
models. In our previous work, we  presented model light curves
in the optical bands with reasonable fits to observations. In this
work, we focus on the near-infrared wavelength region and model the
light curves in the I, J, H, and K bands.

To compute SN Ia light curves, our approach is a simple energy solver that solves for an energy change
of the material during a time step of the envelope evolution. The energy can change due to the
free adiabatic expansion, the deposition of energy from $\gamma$-rays from the radioactive decay
of $^{56}$Ni and $^{56}$Co, and the interaction of radiation with the material.
In our calculations, we assume homologous expansion. The $\gamma$-ray deposition is solved with
the assumption of a gray atmosphere. Changes of ionization state are also included in our solver.

To determine the energy change of the material due to the interaction with the radiation, 
we have to solve the radiative transfer. For each time step, we solve the non-gray spherical symmetric
radiative transfer equation for expanding atmospheres including special relativistic effects.
We typically use about 2,000 wavelength points to keep the computation time reasonable.

Given the problems with the simple LTE treatment of line scattering as shown
in \citet{jack11},
we investigate the effects of IR line
scattering 
on SN~Ia light curves in the near-infrared.
In our
previous paper \citep{jack11}, we  indicated how important UV/optical
line scattering is during the later phases of the light curve
evolution. 
Here, we study this effect in more detail.

The source function of the radiative transfer equation
including scattering for an equivalent two level atom can be written as
\begin{equation}
S_{\lambda}=(1-\epsilon_{\lambda})J_{\lambda}+\epsilon_{\lambda}B_{\lambda}.
\end{equation}
The source function is represented by $S_{\lambda}$, the Planck
function by $B_{\lambda}$ and the mean intensity by $J_{\lambda}$. All
these quantities depend on the wavelength $\lambda$. Here, we used
$\epsilon_{\lambda}$ as the line thermal coupling parameter.  For
$\epsilon_{\lambda}=1$ there is only true absorption and no line
scattering takes place. The source function is then given by
$S_{\lambda}=B_{\lambda}$ (pure LTE). In reality, the thermal coupling
parameter $\epsilon_{\lambda}$ will vary over the whole wavelength
range. For the calculation with \phx, it is possible to set a
wavelength independent factor $\epsilon=\epsilon_{\lambda}=$ constant
to approximate LTE line scattering over the whole wavelength range.

\begin{figure}
\centering
\resizebox{\hsize}{!}{\includegraphics{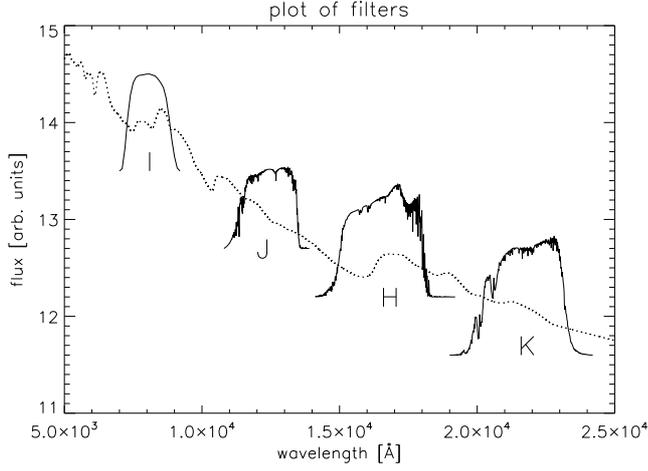}}
  \caption{Synthetic spectrum of a SN Ia at day 20 plotted with the
    filter functions of the I, J, H, and K band.} 
  \label{fig:filters}
\end{figure}
For our approach to model light curves of SNe Ia, we use the results
of hydrodynamical explosion simulations, 
as our initial atmosphere structures which we then evolve in time.
In this work, we use the explosion model results of the deflagration
model W7 \citep{nomoto84}. 
We also used one delayed detonation model, the model DD25 calculated
by \citet{HGFS99by02}. 
Model DD16 \citep {HGFS99by02} which we studied in previous work is
not considered here, because its low yield of $^{56}$Ni makes it too
dim to account for normal SNe~Ia.
We used our hydrodynamic solver to obtain detailed spectra at
certain days in the light curve evolution, \citep[see][for
details]{jack11}. 
Light curves are calculated by convolving the appropriate filter
functions with the theoretical spectra. We compare our synthetic light
curves 
to observed light curves of two SN Ia events: SN~1999ee and
SN~2002bo. The filter functions we used here are shown in 
Fig.~\ref{fig:filters}.
The photometric observations of SN 1999ee in the J and H band are
presented in \citet{krisciunas04}. 
For SN 2002bo photometric observations in the J, H, and K band have
been obtained by \citet{krisciunas04a}.

\section{Near-infrared light curves of SNe Ia}
\label{sec:line_scat}

\begin{figure}
\centering
\resizebox{\hsize}{!}{\includegraphics{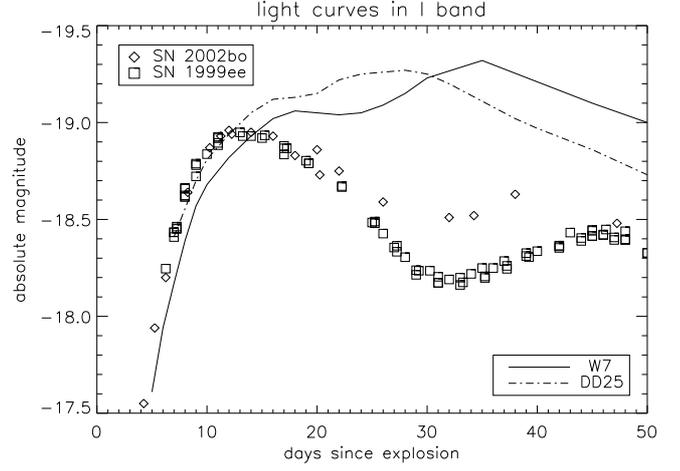}}
  \caption{Model and observed light curves in the I band. Two
    different explosion models were used to compute the model light
    curves.} 
  \label{fig:lc_hyd_Iband}
\end{figure}
We first present the results of an approach of a constant
line scattering factor $\epsilon=0.8$ to the light curve modeling,
where we compare the results of two different hydrodynamical models.
Figure \ref{fig:lc_hyd_Iband} shows the calculated model light curves
of two hydrodynamical models in the I band. They are compared to two observed
light 
curves of SN 1999ee and SN 2002bo. Both model light curves produce
the steep rise during the first days after the explosion reasonably
well. 
However, the model light curve rises further, although the observed
light curves show a decline after maximum at around 13 days 
after explosion. 
During the later phase around 30 days after the explosion, the model
light curves are much too bright compared to the observed light curves.
There is also no indication of a secondary maximum in the model light curves.
\begin{figure}
\centering
\resizebox{\hsize}{!}{\includegraphics{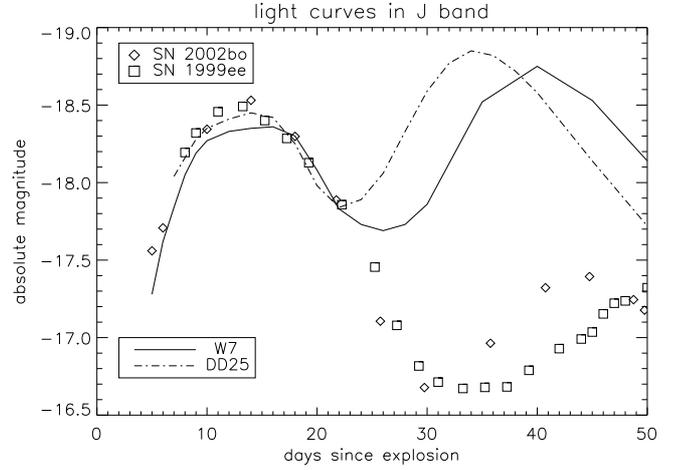}}
  \caption{Light curves in the J band. Two different explosion models
    were used to compute the model light curves and to compare them 
  to observations.}
  \label{fig:lc_hyd_Jband}
\end{figure}
In Fig. \ref{fig:lc_hyd_Jband}, the model light curves in the J band
are shown together with the observed light curves. Again, 
the first phase is well represented by the model light curves. The
model light curves rise after a first maximum to a secondary maximum, 
which is much brighter and earlier than the secondary maximum in the
observed light curves. Furthermore, 
the secondary maximum of the DD25 model is earlier than the one of
hydrodynamical model W7. 
However, the difference of $2.5$ mag around day 35 between models and
observations is enormous.  
\begin{figure}
\centering
\resizebox{\hsize}{!}{\includegraphics{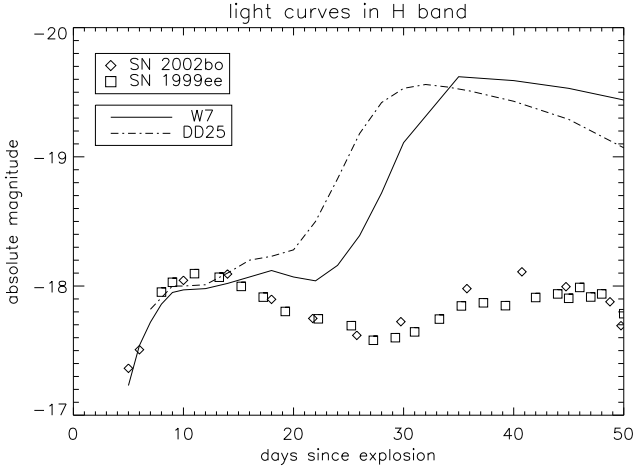}}
  \caption{Light curves in the H band. Two different explosion models
    were used to compute the model light curves.} 
  \label{fig:lc_hyd_Hband}
\end{figure}
\begin{figure}
\centering
\resizebox{\hsize}{!}{\includegraphics{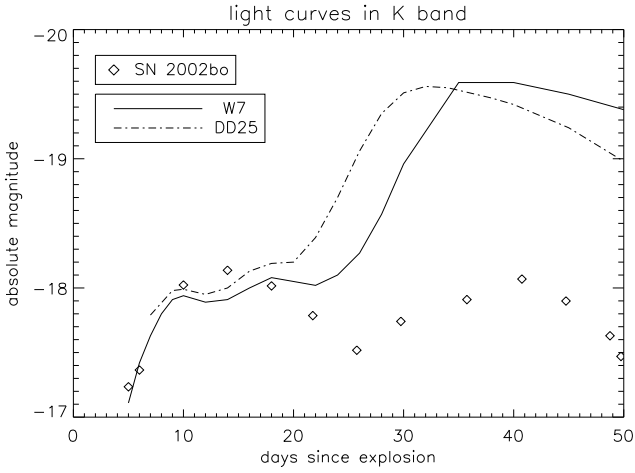}}
  \caption{Light curves in the K band. The model light curves of W7
    and DD25 are significantly brighter than the observations 
   during the later phase.} 
  \label{fig:lc_hyd_Kband}
\end{figure}
A similar picture is seen in the H and K band as shown in Fig.
\ref{fig:lc_hyd_Hband} and Fig. \ref{fig:lc_hyd_Kband}.

So far, we used for the calculations of our LTE model light curves a
constant time independent factor of $\epsilon=0.8$.
We now use this LTE line scattering factor to approximate
the effect of line scattering on the shape of the light curves in the
near-infrared.
We calculated several model light curves each with a different LTE
line scattering factor, $\epsilon$. At each point of the light curve,
we let the atmosphere structure adapt to the new conditions with its
$\epsilon$ until a radiative equilibrium state is reached.  This led
to different resulting atmosphere structures or more precisely
temperature structures at each day and thus to differences in the
shapes of the model light curves.

\begin{figure}
\centering
\resizebox{\hsize}{!}{\includegraphics{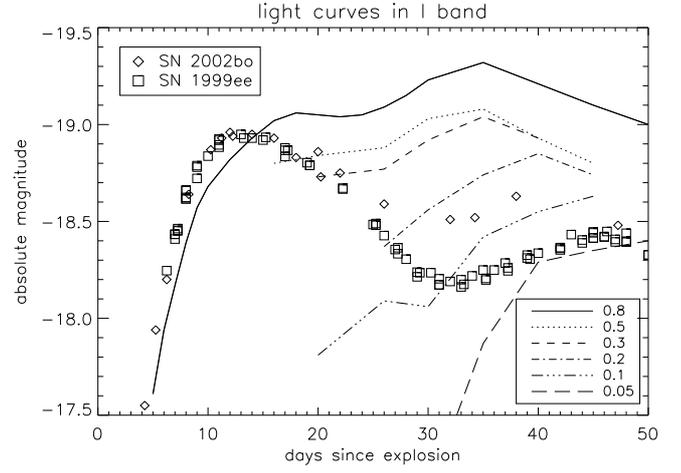}}
  \caption{Light curves in the I band computed with different values
    of the line scattering factor $\epsilon$. } 
  \label{fig:lc_Iband_epslin}
\end{figure}
\begin{figure}
\centering
\resizebox{\hsize}{!}{\includegraphics{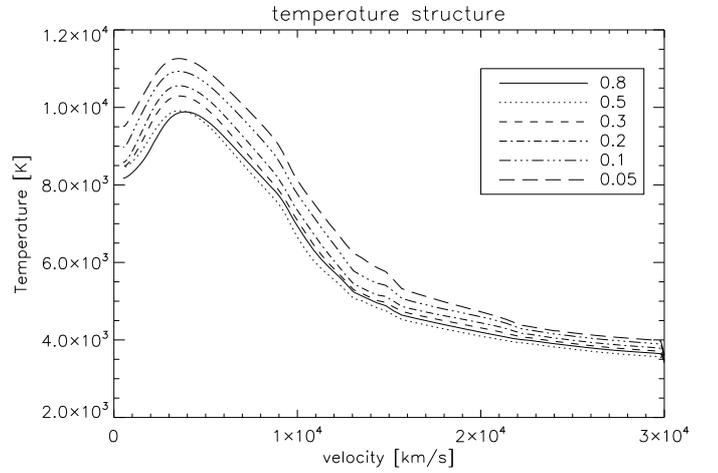}}
  \caption{Temperature structure of the envelope at day 30. A smaller line scattering
    factor $\epsilon$ leads to a hotter atmosphere.} 
  \label{fig:lc_temp_epslin}
\end{figure}
Figure~\ref{fig:lc_Iband_epslin} shows the
I band light curves obtained with different
values of $\epsilon$. With
a smaller $\epsilon$, the light curve becomes fainter. A look at the
temperature structure in Fig. \ref{fig:lc_temp_epslin} reveals that 
decreased $\epsilon$ leads to a hotter atmosphere. 
In the hotter
atmosphere, the iron and cobalt are mostly doubly ionized.
As the atmosphere cools down, the
iron and cobalt recombine to Fe~II and Co~II, and the flux increases
due to numerous lines of these species in the near-infrared. This
recombination effect in the near-infrared light curve has also been
found by \citet{kasen06b} and \citet{hofdd+mol95}. The intersection of the model light curves
with the observed light curves shows that the best value of $\epsilon$
changes during the evolution of the light curve. 
$\epsilon$ decreases with time (i.e., line scattering becomes more and
more important)  
in order to fit the observed light curves.
This is reasonable, because the ongoing
expansion leads to lower densities at later times, and line scattering
becomes more important because of the thinner atmosphere.
However, during the early phase near maximum light, $\epsilon$ has to be
closer to pure LTE \citep{snefe296}.

We now seek to fit the entire light curve with a single function for
$\epsilon$, that is we choose $\epsilon = \epsilon_0 f(t)$ where
$f(t)$ is a function of time.
Our model light curves now reproduce the observed light curves much
more faithfully.
Even the secondary maximum is well reproduced for each
band.
For instance, we show in
Figure \ref{fig:lc_Iband_nofecore} the new best fitting light curve in the I band.
The solid line shows the W7 model light curve, which has a
an $\epsilon$ that decreases with time.
We chose the light curve of SN 1999ee to obtain the best fit. It is clear that
different values of $f(t)$ can be used to obtain a better fit to
the light curve of SN 2002bo.

\section{Secondary maximum in the near-infrared SN Ia light curves} 

In all bands in the near-infrared, the observed light curves show a
secondary maximum around 30 to 40 days after  explosion. Since
with a time-dependent $\epsilon$ our 
model light curves reproduce the observed light curves
pretty well, we will focus on the specific cause of the secondary
maximum in each IR band.
We investigate the cause
of the secondary maximum in each of the near-infrared bands by looking
at the spectral evolution of the model light curves in the respective
band.  Basically, we confirm the results of \citet{kasen06b}. However,
we can additionally assign features of certain elements to the
individual bands and show detailed spectra.

The I band is known to be dominated by the Ca IR triplet.
\begin{figure}
\centering
\resizebox{\hsize}{!}{\includegraphics{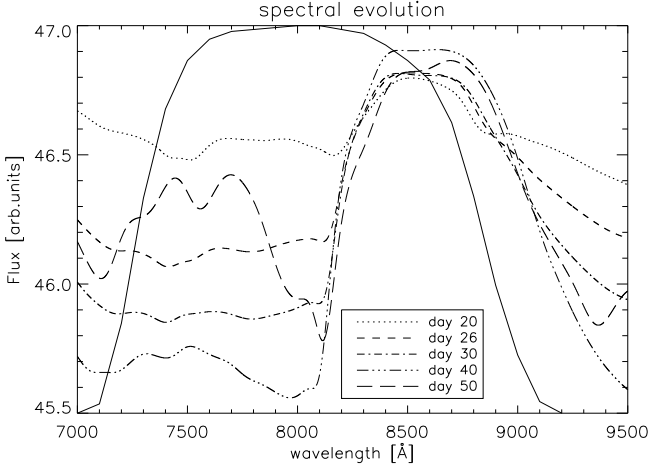}}
  \caption{Spectral evolution in the I band from day 20 to day 50
    after the explosion. The solid line displays the filter
    function.} 
  \label{fig:spec_Iband}
\end{figure}
In Fig. \ref{fig:spec_Iband} the spectral evolution in the I band is
shown. The broad feature at 8500~\AA\ is the Ca triplet.  This feature
stays present as the flux outside of the feature declines while
the SN Ia envelope evolves. This leads to the decline in the light
curve after the first maximum.  Starting at day 40 a new ``W-shaped'' 
feature emerges at around 7500~\AA. These are lines of Fe II, and the
feature is even more significant in the spectrum at day 50.  From day
30 to day 50 the temperature in the expanding envelope is
decreasing. The temperature eventually reaches the temperature/pressure
regime where  Fe~III recombines to Fe~II. This increases the
brightness in the I band due to \emph{emission} lines of Fe~II present
in this wavelength region.  Thus, emission lines of Fe~II are
responsible for the rise to a secondary maximum in the
I~band.

\begin{figure}
\centering
\resizebox{\hsize}{!}{\includegraphics{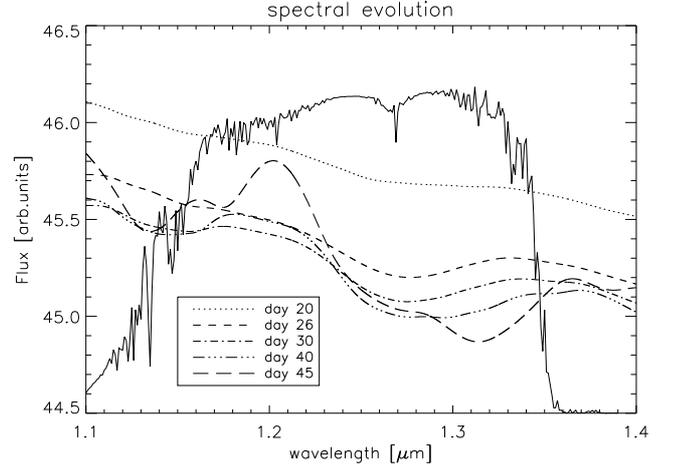}}
  \caption{Spectral evolution in the J band from day 20 to day 45
    after the explosion. The solid line shows the filter
    function.} 
  \label{fig:spec_Jband}
\end{figure}
We now examine the secondary maximum in the J~band. The spectral
evolution in that wavelength range is shown in
Fig. \ref{fig:spec_Jband}.  With the SN Ia evolving, the flux is
decreasing in the J~band because the temperature of the envelope
decreases. At day 30, new features arise, which increase the
brightness in the J~band. These features are caused by lines of Fe~II
and Co~II. Therefore, the recombination of Fe~III to Fe~II is
responsible for the rise in the J~band.  Additionally, Co~III
recombines to Co~II at about the same temperature and therefore time
in the light curve evolution.  Thus, Co~II lines emerge in the J band
due to this recombination.  The secondary maximum in the J band is
caused by both emerging Fe~II and Co~II lines.

\begin{figure}
\centering
\resizebox{\hsize}{!}{\includegraphics{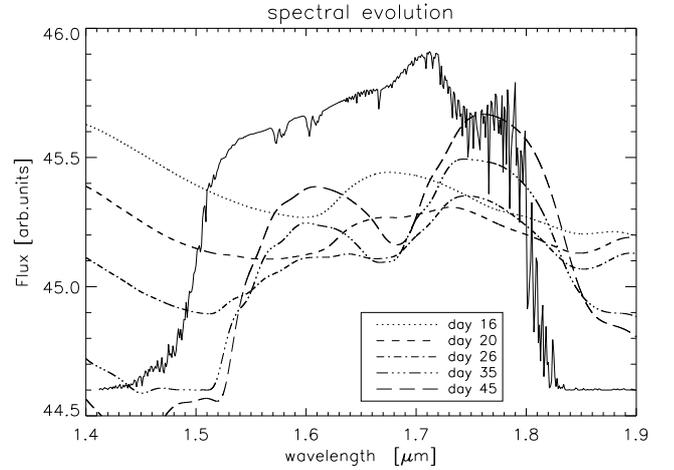}}
  \caption{Spectral evolution in the H band from day 16 to day 45
    after the explosion. The solid line shows the filter
    function.} 
  \label{fig:spec_Hband}
\end{figure}
The spectral evolution in the H~band is shown in
Fig. \ref{fig:spec_Hband}. In the spectrum at day 26, two broad
features emerge in this wavelength range. In the spectrum of day 35
and day 45, these two features are clearly visible. They cause the
rise in the brightness of the H band to a secondary maximum. We found
that these features are caused by lines of Co~II.  Thus, the
recombination of Co~III to Co~II causes the rise to a secondary
maximum in the light curve of the H~band.

\begin{figure}
\centering
\resizebox{\hsize}{!}{\includegraphics{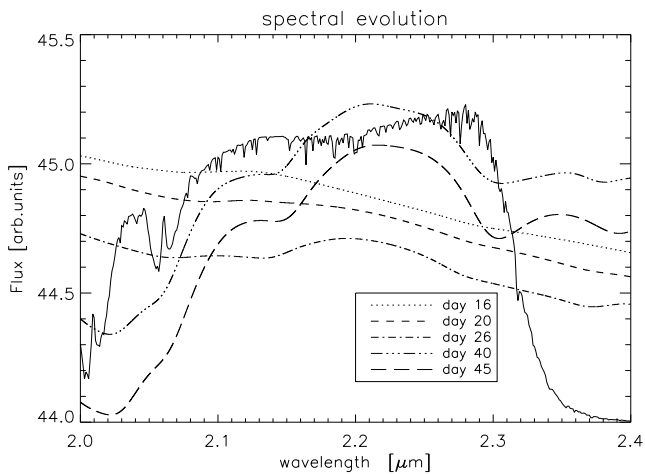}}
  \caption{Spectral evolution in the K band from day 16 to day 45
    after the explosion. The solid line shows the filter
    function.} 
  \label{fig:spec_Kband}
\end{figure}
In the K band, the spectral evolution of our SN Ia model is shown in
Fig. \ref{fig:spec_Kband}.  The flux is decreasing as the envelope
expands and cools down. However, at day 40 and day 45 features emerge
and cause a rise in the brightness of the K band. These features are
again lines of Co~II. The recombination of Co~III to Co~II causes
the secondary maximum in the K~band.

\subsection{Non-radioactive core}

\begin{figure}
\centering
\resizebox{\hsize}{!}{\includegraphics{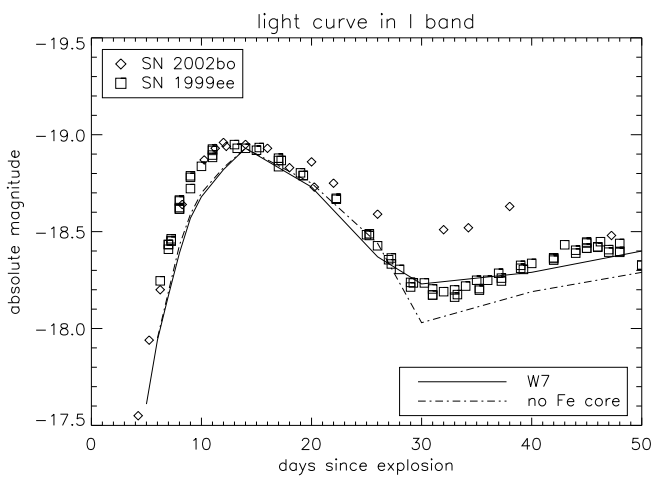}}
  \caption{Light curve in the I band with and without the non-radioactive
  iron core.} 
  \label{fig:lc_Iband_nofecore}
\end{figure}

The W7 model has a non-radioactive iron core. \citet{PE00} showed that
the consequent negative temperature gradient leads to a further reduction 
in the mean opacity. 
This may have an influence on the duration or magnitude of the IR secondary maximum.
Therefore, it is worthwhile checking wether the non-radioactive center affects the shape
of the secondary maximum in the near-infrared.
We replaced the iron core of the W7 model with radioactive $^{56}$Ni and
computed the light curve.  This increases the total mass
of $^{56}$Ni from $0.56$ to $0.66$ solar masses. As shown in Fig. \ref{fig:lc_Iband_nofecore},
there are only small difference between the light curves with and without the
non-radioactive iron core. 
This simple variation does not
capture the full effect of the presence or absence of a
non-radioactive core, since a shift in the nickel distribution can
have important effects on the light curve (E.~Baron {et~al.}, in
preparation). Additionally, late time spectra show that that a
non-radioactive core seems to be required in most SNe~Ia
\citep{hof03du04,moto06,fesen07,gerardy07,maedanature10}.

\section{Conclusion}

We applied our time-dependent model atmosphere code \phx\ to model
light curves of type Ia supernovae in the near-infrared wavelength
range. In a first approach with a constant LTE line scattering
parameter $\epsilon$ in the radiative transfer, we reproduced the
observed light curves during the first phase quite well. However, the
model light curves during the later phase were too bright, and we could
not reproduce the secondary maximum.

We show that a more detailed treatment of IR line scattering is very
important for the modeling of the later phase of the near-infrared
light curves of SNe Ia. We used as an approximation an LTE line
scattering parameter that decreases in time. This is a good approach,
because the atmosphere becomes thinner as expansion goes on and
scattering becomes more important. We use this approximation to
obtain fits to the observed light curves of SN 1999ee and SN 2002bo,
improving our model light curves significantly. The next step is to treat the
atmosphere in full non-LTE, where the temperature needs to adapt to
the non-LTE condition, which requires substantial time on parallel
supercomputers. \citet{kasen06b} also found that his models overshot
the observed secondary maximum in $H$ and $K$, while he mainly attributed this
to incomplete line lists in the IR, he also suggested that his
treatment of LTE might be a possible cause. 

The secondary maximum in each band of the near-infrared was quite
accurately reproduced by our model light curves with time-dependent $\epsilon$.
We investigated the spectral evolution and found that the I band
secondary maximum arises due to recombination of Fe~III to Fe~II.  An
Fe~II feature emerges in this band and increases the brightness.  For
the secondary maximum in the J band, a mix of Fe~II and Co~II lines
emerges and causes the rise to a secondary maximum in the model light
curve.  The bands H and K also show a secondary maximum. This is
caused by the recombination of Co~III to Co~II.  We confirmed that
ionization stage changes in iron group elements are responsible for
the secondary maximum as previously found \citep{kasen06b}. In fact, since we
use neither expansion opacities, nor the Sobolev approximation, our
results robustly indicate the need for a full NLTE treatment.
Furthermore, we
can explicitly assign lines of different elements to different
bands.  These results also show how important the treatment of Fe and
Co in non-LTE is and should be used in future work, when faster
computers are available.

\begin{acknowledgements}
This work was supported in part by the
Deutsche Forschungsgemeinschaft (DFG) via the SFB 676, NSF grant AST-0707704,
US DOE Grant DE-FG02-07ER41517, and by  
program number HST-GO-12298.05-A which is supported by NASA through 
a grant from the Space Telescope Science Institute, which is operated by the 
Association of Universities for Research in Astronomy, Incorporated, under 
NASA contract NAS5-26555.
This research used resources of the National Energy Research
Scientific Computing Center (NERSC), which is supported by the Office
of Science of the U.S. Department of Energy under Contract
No. DE-AC02-05CH11231, and the H\"ochstleistungs Rechenzentrum Nord
(HLRN). We thank all these institutions for generous allocations of
computation time.
\end{acknowledgements}

\bibliographystyle{aa}
\bibliography{17271bib}

\begin{thebibliography}{18}
\expandafter\ifx\csname natexlab\endcsname\relax\def\natexlab#1{#1}\fi

\bibitem[{Baron {et~al.}(1996)Baron, Hauschildt, Nugent, \& Branch}]{snefe296}
Baron, E., Hauschildt, P.~H., Nugent, P., \& Branch, D. 1996, MNRAS, 283, 297

\bibitem[{{Elias} {et~al.}(1981){Elias}, {Frogel}, {Hackwell}, \&
  {Persson}}]{elias81}
{Elias}, J.~H., {Frogel}, J.~A., {Hackwell}, J.~A., \& {Persson}, S.~E. 1981,
  ApJ, 251, L13

\bibitem[{{Fesen} {et~al.}(2007){Fesen}, {H{\"o}flich}, {Hamilton}, {Hammell},
  {Gerardy}, {Khokhlov}, \& {Wheeler}}]{fesen07}
{Fesen}, R.~A., {H{\"o}flich}, P.~A., {Hamilton}, A.~J.~S., {et~al.} 2007, ApJ,
  658, 396

\bibitem[{{Gerardy} {et~al.}(2007)}]{gerardy07}
{Gerardy}, C.~L. {et~al.} 2007, \apj, 661, 995

\bibitem[{H{\"o}flich {et~al.}(2002)H{\"o}flich, Gerardy, Fesen, \&
  Sakai}]{HGFS99by02}
H{\"o}flich, P., Gerardy, C., Fesen, R., \& Sakai, S. 2002, ApJ, 568, 791

\bibitem[{{H{\"o}flich} {et~al.}(2004){H{\"o}flich}, {Gerardy}, {Nomoto},
  {Motohara}, {Fesen}, {Maeda}, {Ohkubo}, \& {Tominaga}}]{hof03du04}
{H{\"o}flich}, P., {Gerardy}, C.~L., {Nomoto}, K., {et~al.} 2004, ApJ, 617,
  1258

\bibitem[{H{\"o}flich {et~al.}(1995)H{\"o}flich, Khokhlov, \&
  Wheeler}]{hofdd+mol95}
H{\"o}flich, P., Khokhlov, A., \& Wheeler, J.~C. 1995, ApJ, 444, 831

\bibitem[{{Jack} {et~al.}(2009){Jack}, {Hauschildt}, \& {Baron}}]{jack09}
{Jack}, D., {Hauschildt}, P.~H., \& {Baron}, E. 2009, \aap, 502, 1043

\bibitem[{{Jack} {et~al.}(2011){Jack}, {Hauschildt}, \& {Baron}}]{jack11}
{Jack}, D., {Hauschildt}, P.~H., \& {Baron}, E. 2011, \aap, 528, A141+

\bibitem[{{Kasen}(2006)}]{kasen06b}
{Kasen}, D. 2006, ApJ, 649, 939

\bibitem[{{Krisciunas} {et~al.}(2004{\natexlab{a}}){Krisciunas}, {Phillips}, \&
  {Suntzeff}}]{KPS04}
{Krisciunas}, K., {Phillips}, M.~M., \& {Suntzeff}, N.~B. 2004{\natexlab{a}},
  \apjl, 602, L81

\bibitem[{{Krisciunas} {et~al.}(2004{\natexlab{b}}){Krisciunas}, {Phillips},
  {Suntzeff}, {Persson}, {Hamuy}, {Antezana}, {Candia}, {Clocchiatti}, {DePoy},
  {Germany}, {Gonzalez}, {Gonzalez}, {Krzeminski}, {Maza}, {Nugent}, {Qiu},
  {Rest}, {Roth}, {Stritzinger}, {Strolger}, {Thompson}, {Williams}, \&
  {Wischnjewsky}}]{krisciunas04}
{Krisciunas}, K., {Phillips}, M.~M., {Suntzeff}, N.~B., {et~al.}
  2004{\natexlab{b}}, \aj, 127, 1664

\bibitem[{{Krisciunas} {et~al.}(2004{\natexlab{c}}){Krisciunas}, {Suntzeff},
  {Phillips}, {Candia}, {Prieto}, {Antezana}, {Chassagne}, {Chen}, {Dickinson},
  {Eisenhardt}, {Espinoza}, {Garnavich}, {Gonz{\'a}lez}, {Harrison}, {Hamuy},
  {Ivanov}, {Krzemi{\'n}ski}, {Kulesa}, {McCarthy}, {Moro-Mart{\'{\i}}n},
  {Muena}, {Noriega-Crespo}, {Persson}, {Pinto}, {Roth}, {Rubenstein},
  {Stanford}, {Stringfellow}, {Zapata}, {Porter}, \&
  {Wischnjewsky}}]{krisciunas04a}
{Krisciunas}, K., {Suntzeff}, N.~B., {Phillips}, M.~M., {et~al.}
  2004{\natexlab{c}}, \aj, 128, 3034

\bibitem[{{Maeda} {et~al.}(2010){Maeda}, {Benetti}, {Stritzinger}, {R{\"o}pke},
  {Folatelli}, {Sollerman}, {Taubenberger}, {Nomoto}, {Leloudas}, {Hamuy},
  {Tanaka}, {Mazzali}, \& {Elias-Rosa}}]{maedanature10}
{Maeda}, K., {Benetti}, S., {Stritzinger}, M., {et~al.} 2010, Nature, 466, 82

\bibitem[{{Motohara} {et~al.}(2006){Motohara}, {Maeda}, {Gerardy}, {Nomoto},
  {Tanaka}, {Tominaga}, {Ohkubo}, {Mazzali}, {Fesen}, {H{\"o}flich}, \&
  {Wheeler}}]{moto06}
{Motohara}, K., {Maeda}, K., {Gerardy}, C.~L., {et~al.} 2006, \apjl, 652, L101

\bibitem[{Nomoto(1984)}]{nomoto84}
Nomoto, K. 1984, ApJ, 277, 791

\bibitem[{{Pinto} \& {Eastman}(2000)}]{PE00}
{Pinto}, P.~A. \& {Eastman}, R.~G. 2000, ApJ, 530, 757

\bibitem[{{Suntzeff}(1996)}]{suntz96}
{Suntzeff}, N.~B. 1996, in IAU Colloq. 145: Supernovae and Supernova Remnants,
  ed. {T.~S.~Kuhn} (Cambridge; UK: Cambridge University Press), 41--+

\end{thebibliography}

\end{document}